\documentclass[aps,twocolumn,groupedaddress,showpacs]{revtex4}
\usepackage{graphicx}
\usepackage{color}
\usepackage{ulem}
\begin{document}
\title{Finite-temperature phase transition to $m=1/2$ plateau phase in a S=1/2 XXZ model on Shastry-Sutherland Lattices}
\author{T. Suzuki}
\affiliation{Institute for Solid State Physics, University of Tokyo, Kashiwa, Chiba 277-8581, Japan}
\author{Y. Tomita}
\affiliation{Institute for Solid State Physics, University of Tokyo, Kashiwa, Chiba 277-8581, Japan}
\author{P. Sengupta}
\affiliation{School of Physical and Mathematical Sciences, Nanyang Technological University, 21 Nanyang Link, Singapore 637371, Singapor}
\author{N. Kawashima}
\affiliation{Institute for Solid State Physics, University of Tokyo, Kashiwa, Chiba 277-8581, Japan}

\date{\today}
\begin{abstract}
We study the finite-temperature transition to the $m=1/2$ magnetization 
plateau in  a model of interacting $S=1/2$ spins with longer range interactions and
strong exchange anisotropy on the 
geometrically frustrated Shastry-Sutherland lattice. This model was shown to capture 
the qualitative features of the field-induced magnetization  plateaus in the rare-earth tetraboride,
${\rm TmB_4}$. Our results show that the transition to the plateau state occurs via two successive 
transitions with the two-dimensional Ising universality class, 
when the quantum exchange interactions
are finite, whereas a single phase transition takes place in the purely Ising limit. To better 
understand these behaviors, we perform Monte 
Carlo simulations of the classical generalized four-state chiral clock model and compare the phase diagrams of 
the two models. Finally, we estimate a parameter set that can explain the magnetization curves observed in 
${\rm TmB_4}$. The magnetic properties and critical behavior of the finite-temperature transition to the 
$m=1/2$ plateau state are also discussed.

\end{abstract}
\pacs{75.40.Mg; 75.10.Hk; 75.40.-s}
\maketitle
\section{INTRODUCTION}
Geometrically frustrated interactions and quantum fluctuation inhibit the stabilization of classical orderings. 
They sometimes become a trigger for the emergence of several exotic orders such as, quantum spin 
ice on the pyrochlore lattice\cite{Onoga}, spin liquid state on the honeycomb lattice\cite{Meng}, 
and multiple magnetization plateaus\cite{Sebastian}. Therefore, quantum spin systems with frustrated interactions 
have attracted great interest from both theoretical and experimental approaches.

The $S=1/2$ antiferromagnetic Heisenberg spin model on the Shastry-Sutherland lattice (SSL)\cite{SSL} is one of 
such systems. The Hamiltonian can be expressed by the nearest neighbor (intradimer) $J$ and the next nearest 
neighbor (interdimer) $J'$ couplings:
\begin{eqnarray}
{\mathcal H}&=&J\sum_{\rm N.N.} {\bf S}_{i}\cdot{\bf S}_{j} + J' \sum_{\rm N.N.N.} {\bf S}_{i}\cdot{\bf S}_{j}.
\end{eqnarray}
Experimentally, $\rm{SrCu(BO_3)_2}$\cite{Kageyama1,Kageyama2} has been 
studied extensively for its realization of the model. In this compound, 
${\rm CuBO_3}$ layers stack along the $c$-axis direction and each magnetic layer
 consists of ${\rm Cu^{2+}}$ ions carrying $S=1/2$ spins 
arranged in an orthogonal dimer structure that is topologically 
equivalent to the SSL. From several experimental observations, it was confirmed 
that the field dependence of the magnetization exhibits multiple 
magnetization plateaus. In these magnetization plateau states, Wigner crystals 
of spin-triplet dimers\cite{Takigawa} are 
realized reflecting the strong competition between the kinetic energy gain and
mutual repulsion of the dimers.

Theoretically, this model has been studied in great detail\cite{Ueda} and several 
states, such as the plaquette singlet state at zero field\cite{Koga,Lauchil} and 
a spin supersolid state in the magnetic fields\cite{Momoi} were predicted. In a 
recent paper by Sebastian et al.\cite{Sebastian}, the possibility of fractional 
magnetic plateaus was discussed in analogy to the quantum Hall effect.

 Fractional magnetic plateaus have recently been discovered at low 
temperatures in rare-earth tetraborides $\rm{RB_4}$ [R is a rare-earth
 element]\cite{TbB4,ErB4,TmB4_1,TmB4_2,TmB4_3,TmB4_4,Siemensmeyer}. The 
magnetic moment carrying  ${\rm R^{3+}}$ ions in these compounds are
arranged in a SSL in the $ab$-plane. In ${\rm TmB_4}$, an extended 
magnetization plateau at $m=1/2$ was confirmed for $H_{c1}\sim 1.9$[T]$<H<H_{c2}\sim$ 
3.6[T] when a magnetic field is applied along the $c$-axis\cite{TmB4_1,TmB4_2}. 
Here $m$ is the normalized value by the saturation magnetization, $m=m_z/m_s$.
In contrast to $\rm{SrCu(BO_3)_2}$ a strong anisotropy 
along the $c$-axis is expected owing to the crystal fields. From specific heat 
measurements\cite{TmB4_2}, it has been suggested that the degeneracy of the $J$=$6$ 
multiplet of ${\rm Tm^{3+}}$ is lifted - the lowest energy state for a single ion 
is the non-Kramers doublet with $J_z$=$\pm 6$ and there exists a large energy gap 
to the first excited doublet. By restricting the local Hilbert space to 
the lowest energy doublet, the low-energy magnetic properties of the material can be
described by a S=1/2 XXZ model with Ising-like exchange anisotropy and a ferromagnetic
transverse coupling as discussed in the next section.

The magnetization curves for the effective Hamiltonian have already been 
calculated\cite{Meng_Wessel,Chang_Yang,Liu}. The effective Hamiltonian is 
the $S=1/2$ Ising-like XXZ model on the SSL and it 
is described by
\begin{eqnarray}
{\mathcal H}&=&\sum_{\rm N.N.} \left( J {\bf S}_{i}\cdot{\bf S}_{j} \right)_{\Delta_\perp}+\sum_{\rm N.N.N.}\left( J'{\bf S}_{i}\cdot{\bf S}_{j} \right)_{\Delta_\perp}\nonumber\\
&-&g\mu_B H\sum_{i}{S_{i}}^{z},
\label{Ham_org}
\end{eqnarray}
where $\Delta_\perp$ denotes the Ising anisotropy and  
$(J_{ij}{\bf S}_{i}\cdot{\bf S}_{j})_{\Delta_\perp}=-|J_{ij}|{\Delta}_{\perp}
\left( S_i^{x}S_j^{x}+S_i^{y}S_j^{y} \right) + J_{ij}S_i^{z}S_j^{z}$.
In the Ising limit $\Delta_\perp =0$, it has been established by several approaches
 - such as Monte Carlo simulations \cite{Meng_Wessel} and tensor 
renormalization-group analysis\cite{Chang_Yang} - that only $m=1/3$ plateau is 
stabilized. The presence of the $m=1/2$ plateau has been confirmed when quantum spin 
fluctuation is included\cite{Meng_Wessel,Liu}, and the ground-state phase diagram for 
$\Delta_\perp >0$ has been calculated. However, even for 
finite $\Delta_\perp$, the $m=1/3$ plateau phase extends over a wider 
range of applied fields than the $m=1/2$ plateau phase. Since the $m=1/3$ plateau 
is not observed in $\rm{TmB_4}$, the above Hamiltonian is insufficient to 
explain the experimental observations in ${\rm TmB_4}$.

\begin{figure}[bth]
  \begin{center}
  \includegraphics[scale=0.6]{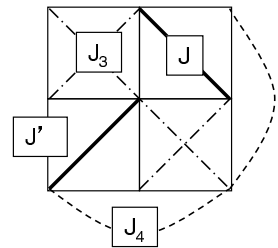}
  \end{center}
  \vspace*{-7mm}
  \caption{Effective model on the SSL with diagonal coupling $J$, nearest-neighbor coupling $J'$, and additional couplings $J_3$ and $J_4$. }
\label{model}
\end{figure}

In a previous letter\cite{Suzuki}, we argued that ferromagnetic $J_4$ and 
antiferromagnetic $J_3$ couplings (see Fig. \ref{model}) are necessary to explain  
the stabilization of an extended $m=1/2$ plateau in the absence of the $m=1/3$ plateau. 
We also investigated the finite-temperature phase transition 
to the $m=1/2$ plateau state. The results of finite-size scaling analysis indicated 
that a two-step second-order transition takes place - the difference between two critical 
temperatures is within 0.5\% of $J$. The universality class at both critical points is 
explained by the critical exponents of the two-dimensional Ising model. 

For the finite-temperature transition, a nontrivial question remains unanswered. 
In the $m=1/2$ plateau phase, the lowest energy state is four-fold degenerate. Therefore, 
it is naively expected that the universality class should be the same as 
that of the four-state Potts model. As we discuss in this paper, the 
critical behavior can indeed be explained by the four-state Potts universality in the 
Ising limit, while the quantum spin model shows a two-step transition with 
both transitions belonging to the two-dimensional Ising universality class. 
Hence the phase diagrams for the thermal phase transitions for
the two are different although both models possess the same symmetry. The 
low energy behavior of the $m=1/2$ plateau  is shown to be described by the 
generalized four-state chiral clock model. As far as we know, the finite-temperature transition 
of the generalized chiral four-state clock model has not been studied precisely. Therefore, 
clarification of the finite-temperature transition to the $m=1/2$ plateau phase is also 
valuable from the view point of statistical mechanics. In the present paper, 
we discuss the properties of the model introduced in our previous letter in greater details, 
and clarify the nature of the phase transitions to the $m=1/2$ plateau state.

This paper is organized as follows. In section II, we reproduce the derivation of an 
effective Hamiltonian that captures the unique magnetization features in  ${\rm TmB_4}$. 
An approach from the Ising limit of the effective Hamiltonian shows that $J_4$ couplings 
are important to stabilize the $m=1/2$ plateau state. In section III, we focus on the 
finite-temperature properties of the effective model and re-investigate the universality 
class of the transition to the $m=1/2$ plateau phase by extensive quantum Monte Carlo 
simulations. We show that the critical properties of the quantum model is different from 
those in the Ising limit. In section IV, a classical model, the generalized chiral four-state 
clock model, is proposed in order to understand the critical behavior of the original model. 
From  Monte Carlo simulations, we find that the topology of the phase diagram for the 
classical model is the same as that of the original model. In comparison with the phase 
diagram for the classical model, we discuss the effects of quantum exchange on the 
finite-temperature transition. In section V, we discuss the magnetic properties of 
${\rm TmB_4}$ and predict the critical properties of the finite-temperature transition 
to the $m=1/2$ plateau phase. Sec. VI is devoted to a summary of the results.

\section{EFFECTIVE MODEL}

In this section, we derive the effective model for the low energy part of magnetic properties of ${\rm TmB_4}$ again.
As discussed in the introduction, the ground state of the single ${\rm Tm^{3+}}$ ions is a $J_z=\pm 6$ doublet with a large energy gap between the first excited doublet. At low temperatures
(compared to this energy gap), one can ignore the contributions from the higher multiplets
and the low energy properties of the material are well described by an effective two-state
model consisting of the $J_z=\pm 6$ doublet. 
The fact that the interactions between the moments are derived from itinerant electrons enables us to expect isotropic type (Heisenberg type) interactions. 
Therefore, we start from the Hamiltonian, $H_{eff}'=H_o+H'$, where $H_o=-D \sum_i (J_i^z)^2$, $H'= \sum_{\langle ij \rangle} G_{ij}{\bf J}_i {\bf J}_j$, and $0<|G_{ij}|\ll D$, where $G_{ij}$ is the coupling constant between two moments on the site $i$ and $j$, and $D$ is the easy-axis anisotropy from the crystal fields.
By treating $H_o$ as the biggest term, we find that there are four degenerate states, namely $|J_i^z,J_j^z\rangle= |6,6\rangle$, $|-6, 6\rangle$, $|6,-6\rangle$, and $|-6, -6\rangle$, in the lowest energy level. 
We apply perturbation theory for $H_{eff}'$ and calculate the matrix elements among these four states. The Hamiltonian for arbitrary moment pairs can be described by the $S=1/2$ Ising-like XXZ Hamiltonian, $h_{ij}=\left( J_{ij} {\bf S}_{i}\cdot{\bf S}_{j} \right)_{\Delta_\perp}$, where $(J_{ij}{\bf S}_{i}\cdot{\bf S}_{j})_{\Delta_\perp}=-|J_{ij}|{\Delta}_{\perp}\left( S_i^{x}S_j^{x}+S_i^{y}S_j^{y} \right) + J_{ij}S_i^{z}S_j^{z}$.
Most importantly, the matrix elements for the transverse coupling ($\langle 6,-6|H'|-6,6\rangle$ and $\langle -6,6|H'|6,-6\rangle$) are proportional to $-\{G_{ij}/(-D)\}^{2J}$. 
Consequently, if the magnetic moment $J$ is even, the transverse coupling always becomes ferromagnetic.
 (Note that the longitudinal component of the interaction becomes ferromagnetic or antiferromagnetic depending on the sign of $G_{ij}$.)

The interaction $G_{ij}$ is expected to be the RKKY type, because this compound 
is a metal. Therefore, the effect of the long-range interactions is significant 
when the magnetic properties of ${\rm TmB_4}$ are considered. The 
principal interactions necessary to reproduce the dominant features of magnetization
in  ${\rm TmB_4}$ - in addition to the conventional SSL model with $J$ and $J'$ 
couplings - are the $J_3$ and $J_4$ couplings, shown in Fig. \ref{model} (f). There 
is another coupling with shorter range than that of $J_3$ and $J_4$ 
- the next-nearest neighbor coupling $J_2$ that is orthogonal to $J$ 
 in the plaquettes with the diagonal interactions in the original SSL 
model. However, as we show in the following analysis, $J_2$ is not efficient in stabilizing 
the $m=1/2$ plateau state, because it stabilizes the $m=1/3$ plateau at the same time.

The Hamiltonian considered here is described by 
\begin{eqnarray}
{\mathcal H}&=&\sum_{\langle i,j \rangle} \left( J {\bf S}_{i}\cdot{\bf S}_{j} \right)_{\Delta_\perp}+\sum_{\langle i,j \rangle'}\left( J'{\bf S}_{i}\cdot{\bf S}_{j} \right)_{\Delta_\perp}\nonumber\\&+&\sum_{\langle i,j \rangle''}\left( J_3{\bf S}_{i}\cdot{\bf S}_{j} \right)_{\Delta_\perp}
+\sum_{\langle i,j \rangle'''}\left( J_4{\bf S}_{i}\cdot{\bf S}_{j} \right)_{\Delta_\perp}\nonumber\\
&-&g\mu_B H\sum_{i}{S_{i}}^{z},
\label{Ham}
\end{eqnarray}
where $\langle ij \rangle$, $\langle ij \rangle'$,  $\langle ij \rangle''$, and 
$\langle ij \rangle'''$ denote sums over all pairs on the bonds 
with the $J$, $J'$, $J_3$, and $J_4$ couplings, respectively. The 
positive (negative) sign of each coupling denotes antiferromagnetic (ferromagnetic) 
interaction. The original SSL model interactions are always assumed to
be antiferromagnetic, i.e., $J>0$ and $J'>0$. 
In the following, we set $J$ as the unit of energy and express all the parameters 
of the model in units of $J$.
We studied the above model on square lattices of the form $L\times L$ with
periodic boundary conditions.

\begin{figure}[bth]
  \begin{center}
  \includegraphics[scale=0.5]{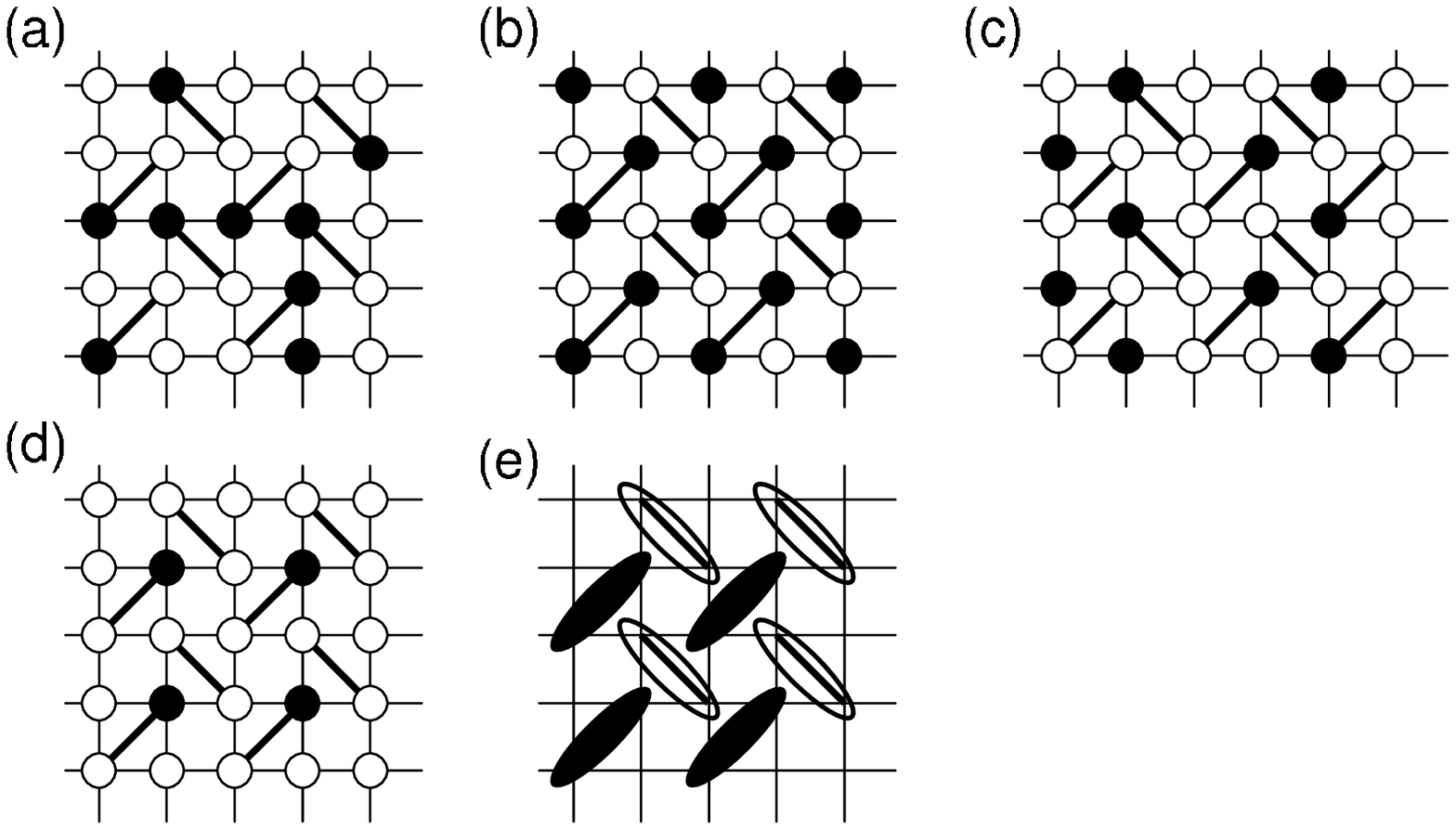}
  \end{center}
  \caption{Conventional spin configurations in (a) an antiferromagnetic classical dimer (ACD) state, (b) the N\'eel state, (c) 1/3 plateau state, and (d) 1/2 plateau state. Solid and open circles indicate down and up spins respectively. (e) Schismatic spin configuration of $C_2$-symmetric phase (see text). Solid (open) ellipses mean spin pairs having the total magnetization $S^z \sim 0$ ($S^z \sim 1$). }
\label{spinconfig}
\end{figure}

\begin{figure}[bth]
  \begin{center}
  \includegraphics[scale=0.45]{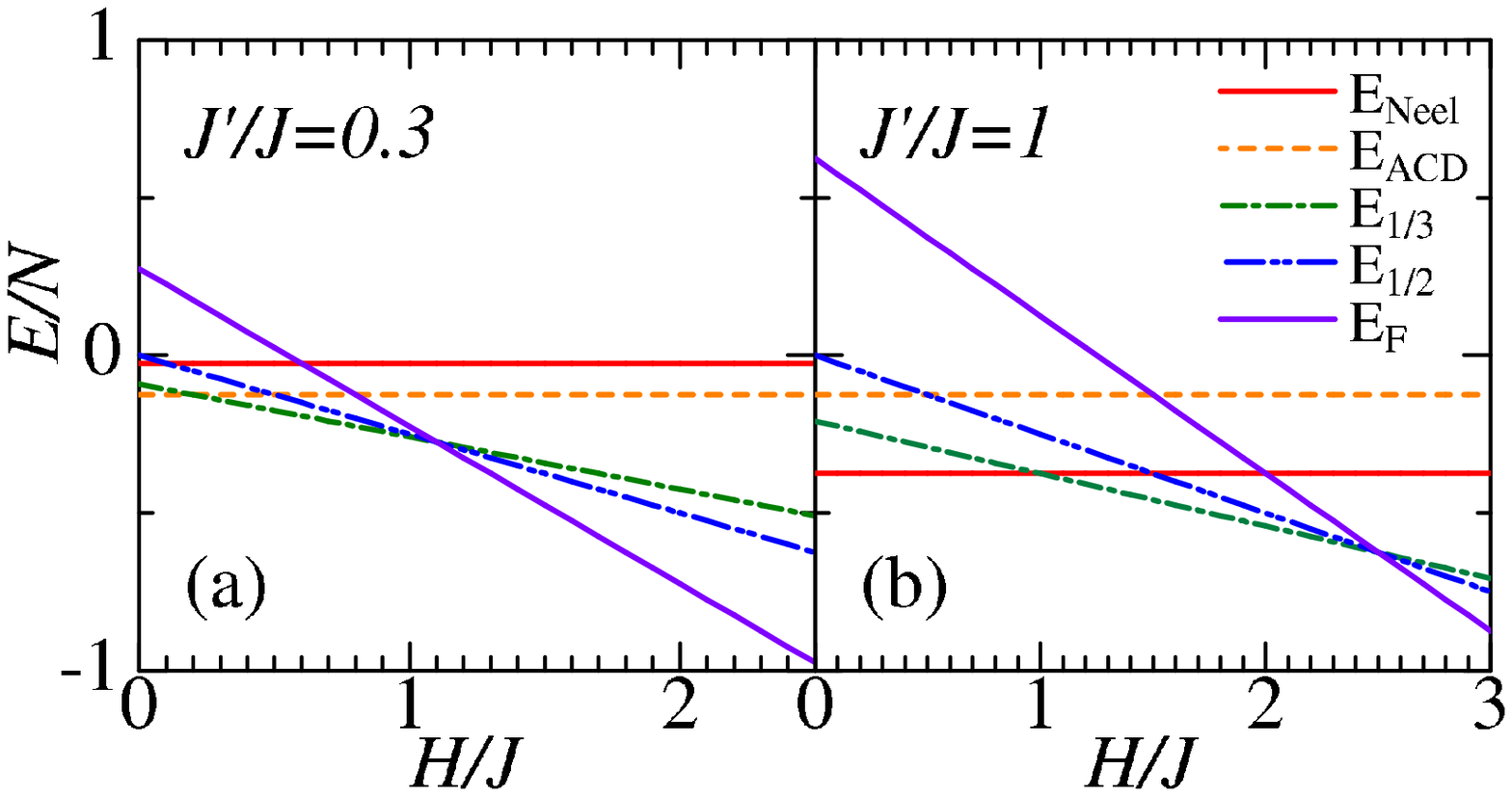}
  \end{center}
  \caption{(color online) Field dependence of the local energy for the Ising limit.}
\label{Isingmodel}
\end{figure}

When $\Delta_{\perp} \ll 1$, the magnetic properties of the Hamiltonian (\ref{Ham}) 
are qualitatively explained by considering the Ising limit ($\Delta_\perp =0$). At 
$J_3=J_4=0$, there are two candidates of the ground states at 
$m=m_z/m_s=2N^{-1}\sum_i {S^z}_{i}=0$, where $N=L^d$ and $d=2$.
For $J'/J<0.5$, each spin pair located on the diagonal bond $J$ forms an 
antiferromagnetic $classical$ dimer (ACD) (Fig. \ref{spinconfig} (a)), but macroscopic 
degeneracy remains due to cancellations among interactions on the coupling 
$J'$.
The local energy of such state can be estimated as $E_{ACD}=-J/8$. In the 
other limit $J'/J \gg 1$, the system approaches the antiferromagnetic 
Ising model on the square lattice, and the N\'eel state becomes the ground state  
(Fig. \ref{spinconfig} (b)). The local energy of the N\'eel state is calculated as 
$E_{Neel}=J/8-J'/2$. As is expected, the energy levels of the two states cross at 
$J'/J=0.5$ and $T=0$, and the system shows a phase transition at the point. The local 
energies of the $1/2$ and $1/3$ plateau state are estimated in the same manner. The spin 
configurations in Fig. \ref{spinconfig} (c) and (d) are realized in the $1/3$ and $1/2$ 
plateau phase, respectively. (These configurations were also confirmed from the snapshot 
of spin configuration in the Monte Carlo simulations in our previous study\cite{Suzuki}.) 
From the spin configurations, we obtain $E_{1/3}=-J/24-J'/6-H/6$ and $E_{1/2}=-H/4$, 
where $E_{1/3}$ and $E_{1/2}$ denote the local energy of the $1/3$ and $1/2$ plateau 
state. Since the energy of the fully polarized state is given by $E_{F}=J/8+J'/2-H/2$, 
the magnetization curve shows a jump from the $m=1/3$ to $m=1$ at $H_c=J/2+2J'$.
We show the field dependence of local energy for these five states in Fig. \ref{Isingmodel}.
The energies of the $m=1/2$ and $m=1/3$ plateaus, and the fully polarized 
state are always degenerate at the saturation field $H_c$. Consequently, the degeneracy between these three 
states should be lifted when the additional couplings $J_3$ and $J_4$ are included. 
By estimating the energy gains due to $J_3$ and $J_4$, we find that the ferromagnetic 
$J_4$ coupling is one of the most efficient ways to stabilize the $1/2$ plateau 
\cite{Footnote}. This analysis further demonstrates that the coupling $J_2$, 
which is perpendicular to the diagonal coupling $J$, does not stabilize the $m=1/2$ plateau, 
because the number of antiparallel spin pairs on the $J_2$ couplings equals that of the 
parallel ones. Hence this coupling was not included in the effective Hamiltonian.

\section{Quantum Monte Carlo Simulations}

The qualitative discussion of the previous section is supplemented
by large scale numerical simulation of the underlying model. We developed
a variant of the standard Stochastic-Series-Expansion quantum-Monte-Carlo method based
on the modified directed-loop algorithm to treat the longer-range interactions efficiently.
\cite{KatoYasu,Todo} Using this new algorithm, simulations of the Hamiltonian
 (\ref{Ham}) were performed on square lattices $L\times L$. Note that the negative sign 
problem does not appear in the quantum Monte Carlo computations for the Hamiltonian (\ref{Ham}), 
because the transverse coupling is ferromagnetic as discussed in the previous section.

\begin{figure}[bth]
  \begin{center}
  \includegraphics[scale=0.45]{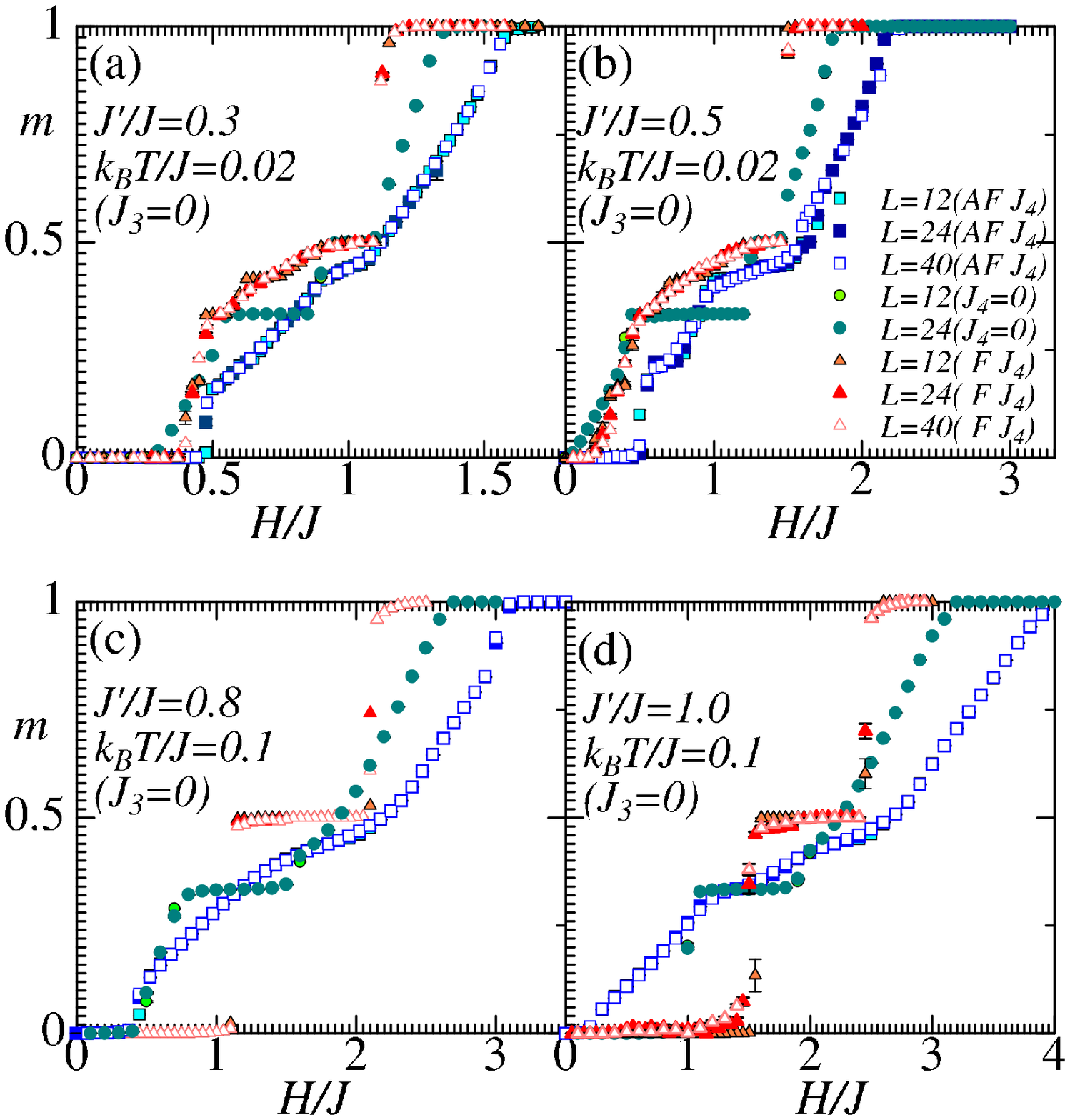}
  \end{center}
  \caption{Magnetization curves when $J_4/J=(1/\sqrt{2})^{3}J'/J$ (color online) and $\Delta_{\perp}=0.2$. Circles indicate the results at $J_4=0$ case. Triangles and squares are the results of the ferromagnetic and antiferromagnetic longitudinal $J_4$ couplings, respectively. All symbols accompany error bars, which are smaller than the symbol size (here and the following figures).}
\label{magnetization}
\end{figure}

Fig. \ref{magnetization} shows the magnetization curves up to $L=40$, for different 
values of the ratio $J'/J$, and how the curves evolve upon the introduction
of ferro- and antiferro-magnetic $J_4$.
In the computations, we fixed the $J_4$ coupling constant at 
$|J_4/J|=(1/\sqrt{2})^3 \times J'/J \sim 0.106$ and the Ising anisotropy 
$\Delta_{\perp}$ at $\Delta_{\perp}=0.2$.

At $J_3=J_4=0$, two magnetization plateaus appear at $m=1/2$ and 1/3, 
consistent with previous studies \cite{Meng_Wessel,Liu,Chang_Yang}. 
When the longitudinal component of $J_4$ coupling is ferromagnetic, 
the system shows a strong tendency towards the stabilizing an extended plateau at $m=1/2$. 
Snapshots of the Monte Carlo simulations in the $m=1/2$ plateau phase confirm
the realization of spin configuration shown in Fig. \ref{spinconfig} (d). 
As long as there is the finite ferromagnetic $J_4$ coupling, the $m=1/3$ plateau region shrinks, 
for all values of the ratio $J'/J$. 
This follows naturally from the analysis of the Ising spin model discussed above. 
In the Ising limit, the field extent of the $m=1/2$ plateau {\it expands} by an amount 
proportional to 4$|J_4|/J$, while that of the $m=1/3$ 
plateau {\it contracts} by an amount proportional to 6$|J_4|/J$.

\begin{figure}[bth]
  \begin{center}
  \includegraphics[scale=0.45]{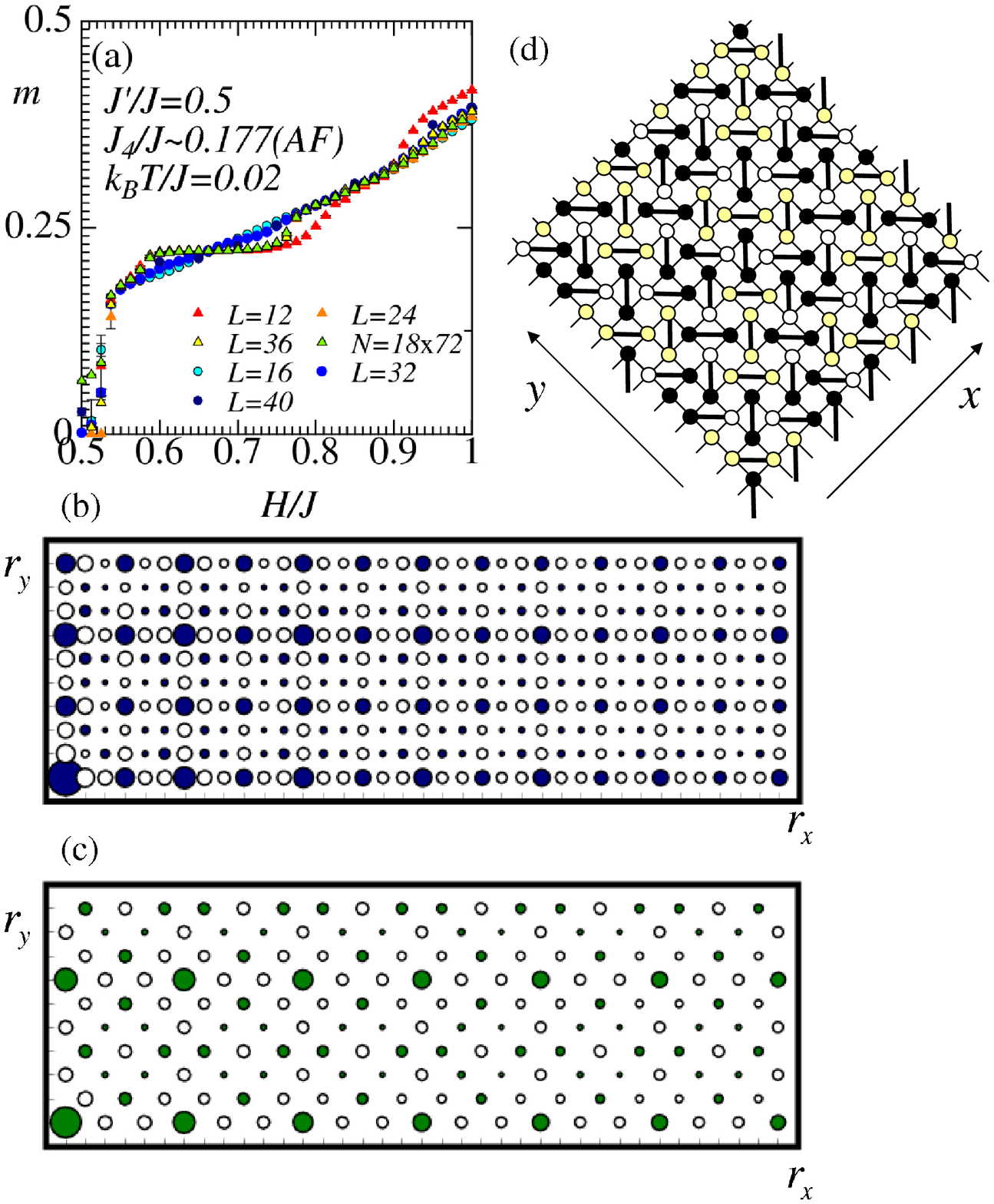}
  \end{center}
  \caption{(color online) (a) Magnetization curves when 
$J'/J=0.5$ and the antiferromagnetic $J_4/J=(1/\sqrt{2})^{3}J'/J$. (b) and (c) 
are results for spin-spin and bond-spin correlations at $k_BT/J=0.02$. Colored 
(open) circles correspond to the positive (negative) amplitude. The area of circles 
is proportional to the amplitude. Error bars are smaller than the symbol sizes.   (d) Schematic 
spin configuration in the $m=2/9$ plateau state. Solid circles denote up spins. Yellow 
circles surrounded by up spins are occupied by up or down spins with almost equal probability.}
\label{2/9plateau}
\end{figure}

When the longitudinal component of $J_4$ coupling is antiferromagnetic  both 
$m=1/2$ and $1/3$ plateaus disappear. 
For antiferromagnetic $J_4$, no conclusive evidence of any plateaus has been obtained 
in our calculations except for $J'/J=0.5$. 
A feature appears in the magnetization curve around $m=1/4$ for $J'/J=0.5$. 
Fig. \ref{2/9plateau} (a) shows the expansion of the magnetization curve 
around $H\sim 0.75$ for $J'/J=0.5$. For $0.60<H/J<0.74$, the magnetization curve becomes flat 
at $m=2/9$. Significantly, the $m=2/9$ plateau was also obtained for the 
parent Shastry-Sutherland model and observed in ${\rm SrCu(BO_3)_2}$.\cite{Sebastian}
To identify the spin structure, we calculated the spin-spin and the 
bond-spin correlation functions. (As for the definition of bond spins, please see 
eq. (\ref{bond-spin})) The results are shown in Fig. \ref{2/9plateau} (b) and (c), 
respectively. In this plateau region, the $3\sqrt{2} \times 3\sqrt{2}$ structure 
accompanying $90\,^{\circ}$ rotational symmetry breaking is clearly stabilized. 
This is the same symmetry breaking as the $m=1/2$ plateau state.
This plateau state has a characteristic feature: the periodicity of the $m=2/9$ plateau state 
is longer than the distance of $J_4$ couplings. 
Furthermore, the periodicity of this plateau is the same as that discussed in ref. \cite{Mila}, but the precise spin configuration is a little bit different from their results shown in figure 5 in ref. \cite{Mila} because the magnitude of the moments on the qlaquette without the diagonal bonds is lager than that on the diagonal bonds having total $S^z=1$.


In the following, we focus on the finite-temperature transition to the $m=1/2$ plateau 
state. The snapshot of the spin configuration of the $m=1/2$ plateau phase leads us to 
expect the $90\,^{\circ}$ ($C_4$) rotation symmetry breaking around the center of the 
plaquette without diagonal coupling $J$. In context of the bare spin language, the 
universality class of the finite-temperature transition is expected to be the four-state 
Potts universality class because the lowest energy state in the $m=1/2$ plateau is four-fold 
degenerate. However, the other scenario is also possible for $\Delta_{\perp} \ne 0$ because 
the group $C_4$ has a subgroup $C_2$. By introducing the $90\,^{\circ}$ lattice rotation 
``$c_4$'' around the center, we can express the symmetry group in the paramagnetic phase 
as $C_4=\{e,c_4,c_4^2,c_4^{-1}\}$. If the system exhibits a two-step phase transition, 
the symmetry breaks down to $\{e,c_4^2\}$ at the higher critical temperature $T_{c1}$, 
and the remaining symmetry breaks down to the trivial group $\{e\}$ at the lower critical 
temperature $T_{c2}$. 

\begin{figure}[bth]
  \begin{center}
  \includegraphics[scale=0.5]{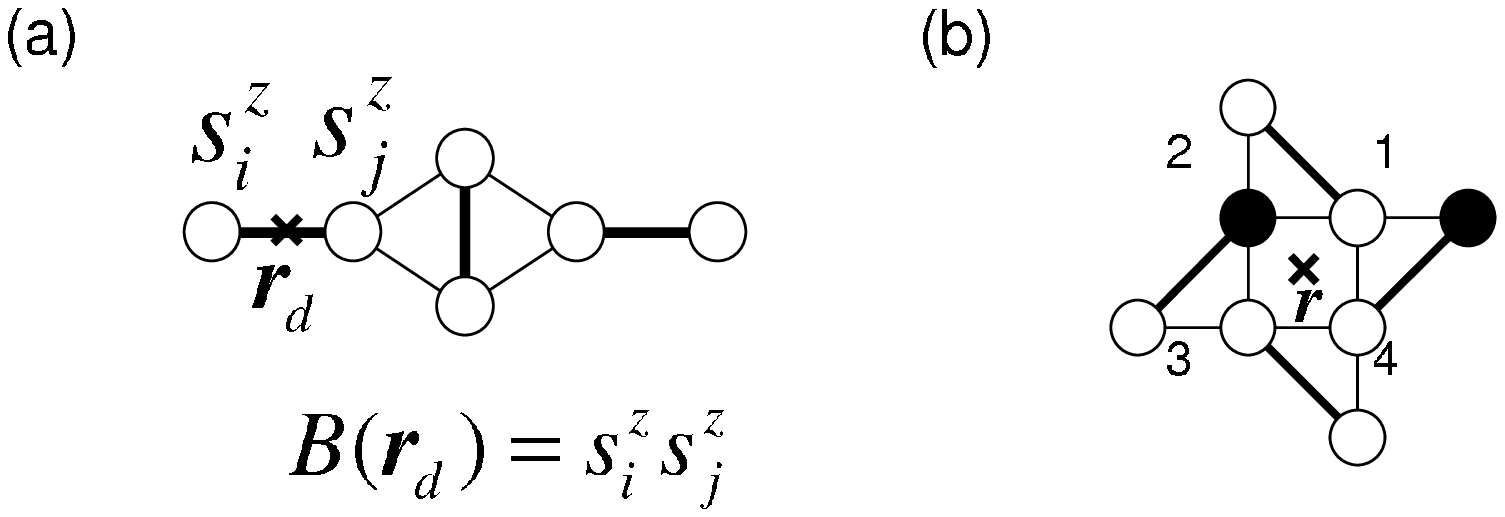}
  \end{center}
  \caption{(a) A bond spin operator. The bond spin operator $B$ is defined by inner product of pair spins on each diagonal bond (bold line) and $B_{st}$ is the staggered magnetization of $B$. The bold black and gray lines indicate the sign of the factor $f({\bf r}_d)$. (b) Definition of sublattice. }
\label{orderparameter}
\end{figure}

To investigate the symmetry breaking at the critical point, we introduce two order parameters. 
The order parameter of $90\,^{\circ}$ rotation symmetry breaking can be expressed by
\begin{eqnarray}
B_{st}=\langle \frac{2}{L^2} |\sum_{{\bf r}_d} (-1)^{f({\bf r}_d)}B({\bf r}_d)|\rangle,
\label{bond-spin}
\end{eqnarray}
where ${\bf r}_d$ is position of the diagonal bond, $B({\bf r}_d)$ is a product of 
longitudinal component of two spins on the bond ${\bf r}_d$, and $f({\bf r}_d)$ takes 
$\pm 1$ depending on the position of the diagonal coupling (Fig. \ref{orderparameter} (a)). 
That of $180\,^{\circ}$ rotation symmetry breaking is also given by
\begin{eqnarray}
 m_x^c&=& \langle \frac{4}{L^2} \sum_{{\bf r}} \{ S^z_4({\rm{\bf r}})-S^z_1({\rm{\bf r}}) \} \rangle,\nonumber\\
 m_y^c&=& \langle \frac{4}{L^2} \sum_{{\bf r}} \{ S^z_3({\rm{\bf r}})-S^z_2({\rm{\bf r}}) \} \rangle,
\end{eqnarray}
where the suffixes of longitudinal spin operators represent the site indexes shown in Fig. \ref{orderparameter} (b), ${\rm{\bf r}}$ is a positional vector for plaquette.  For $T>T_{c1}$, the system is in the paramagnetic phase, and $B_{st}=0$ and $m_x^c=0$ are satisfied. For $T_{c1}>T>T_{c2}$, the system retains the $180\,^{\circ}$ ($C_2$) rotation symmetry, and then $B_{st}\ne0$ and $m_x^c=0$.  Below $T_{c2}$, the lowest energy state is characterized by $B_{st}\ne0$ and $m_x^c\ne0$.

\begin{figure}[bth]
  \begin{center}
  \includegraphics[scale=0.4]{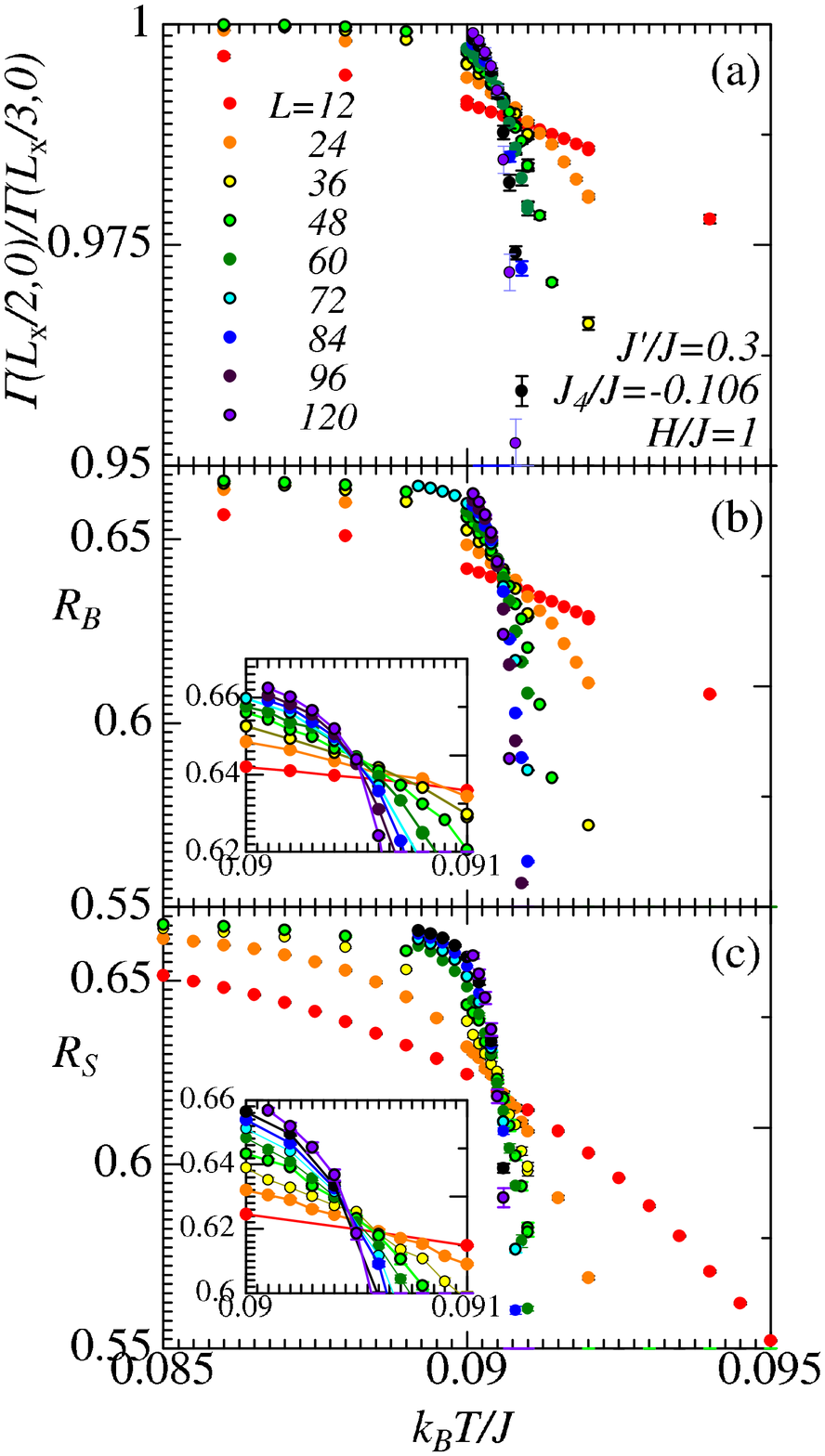}
  \end{center}
  \vspace*{-7mm}                                                        
  \caption{ (color online) Temperature dependence of the correlation ratio at $J_4/J \sim -0.106$ in the Ising limit case. (a) Correlation ratio of the bare spin correlation. (b) The Binder ratio $R_B$ of the bond-spin operator $R_{st}$. (c) The Binder ratio $R_S$ of $m_x^c$.  }
\label{Ising}
\end{figure}

\begin{figure}[bth]
  \begin{center}
  \includegraphics[scale=0.45]{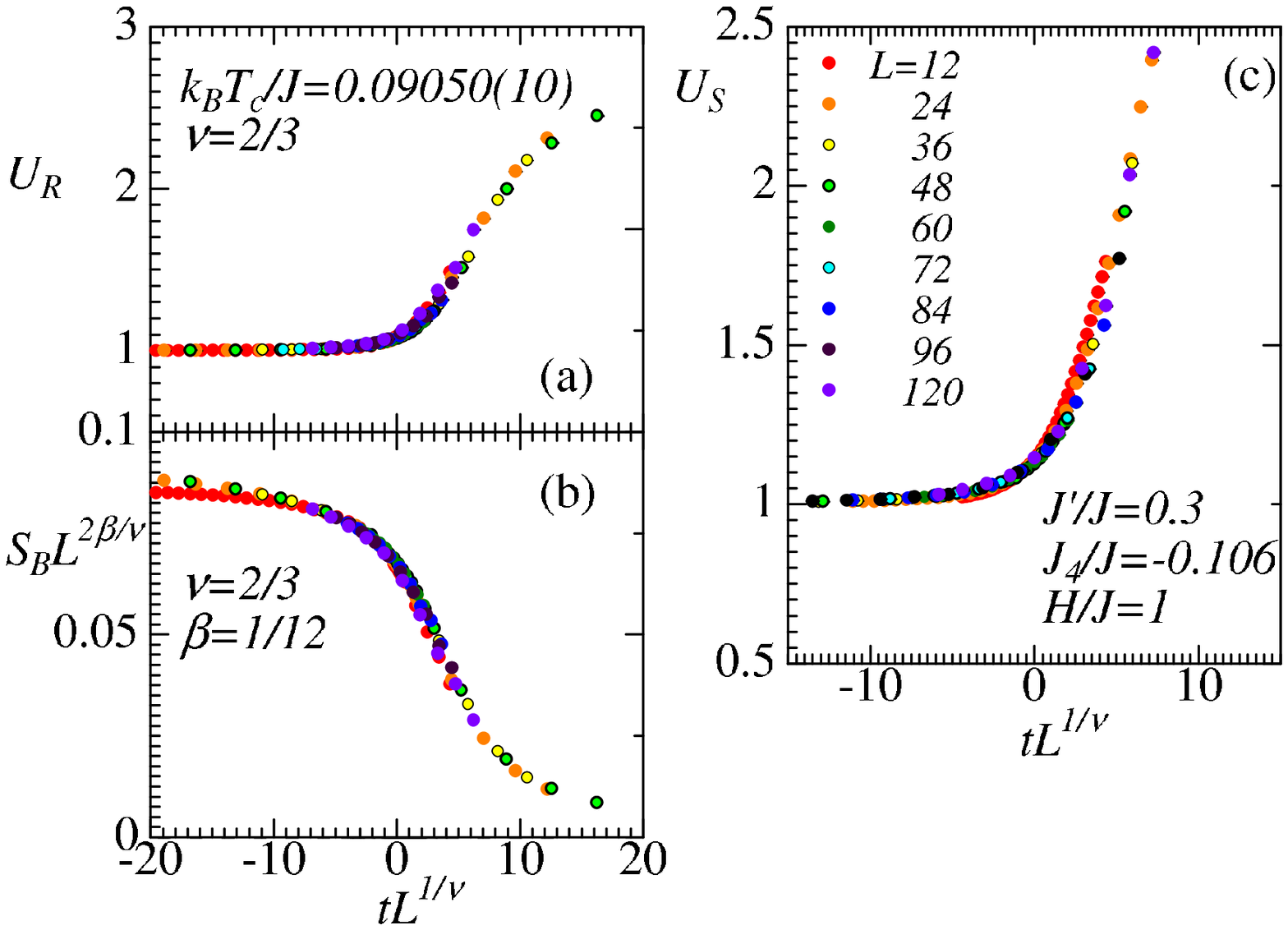}
  \end{center}
  \caption{(color online) Finite-size scaling analysis for $U_R$, $S_B$, and $U_S$ in the Ising limit case. (a) and (c) are the results for the fourth-order cumulant $U_R$ and $U_S$. (b) Scaling analysis for the static structure factor $S_B$ of bond spins.}
\label{IsingFSS}
\end{figure}

First, we show the results of the finite-temperature transition in the Ising limit
($\Delta_{\perp}=0$). The Monte Carlo simulations up to $L=120$ were performed at $J'/J=0.3$ 
and $J_4/J=-(1/\sqrt{2})^3 \times J'/J \sim -0.106$. Fig. \ref{Ising} (a) displays 
the temperature dependence of the correlation ratio of the bare spins at 
$(x,y)=(L_x/2,0)$ and $(L_x/3,0)$ for different system sizes. The 
curves cross at the critical temperature $T_c\sim 0.09050(10)$,  
reflecting the size invariance of the correlation ratio at the critical temperature. 
The spin fluctuations freeze at 
this temperature. Fig. \ref{Ising} (b) and (c) show the temperature dependence of 
the Binder ratios $R_B$ for $B_{st}$ and $R_S$ for $m_x^c$, respectively. The system 
size dependence of both Binder ratios disappears at the same critical temperature. 
The crossing of the curves for $R_B$ and $R_S$ also indicates that the 
$C_4$ symmetry breaks down to the trivial group at $T_c\sim 0.09050(10)$. 
The obtained results suggest that the transition belongs
to the universality class of the four-state Potts model.
This is confirmed by finite-size scaling analysis. For the fourth 
order cumulant of $B_{st}$ and $m_x^c$, we assume the scaling forms $U_R=F(tL^{1/\nu})$ 
and $U_S=F(tL^{1/\nu})$, where $F$ is a scaling function and $U_R$ and $U_S$ 
denote the fourth order cumulant of $B_{st}$ and $m_x^c$, respectively. For the 
static structure factor  of the bond spins, $S_B L^{2\beta/\nu}=F(tL^{1/\nu})$ 
is assumed, where $S_B=\sum_{{\bf r}_d} B({\bf r}_d) \exp{[i\pi(r_x + r_y)]}$. In the 
analysis, the critical exponents and the critical temperature are fixed at $\nu=2/3$, 
$\beta=1/12$\cite{Domb} and $T_c=0.09050(10)$, The results are shown in Fig. 
\ref{IsingFSS}. The excellent data collapse confirms that the transition 
in the Ising limit belongs to the universality class of the four state Potts model.


\begin{figure}[hbt]
  \begin{center}
  \vspace*{2mm}
  \includegraphics[scale=0.45]{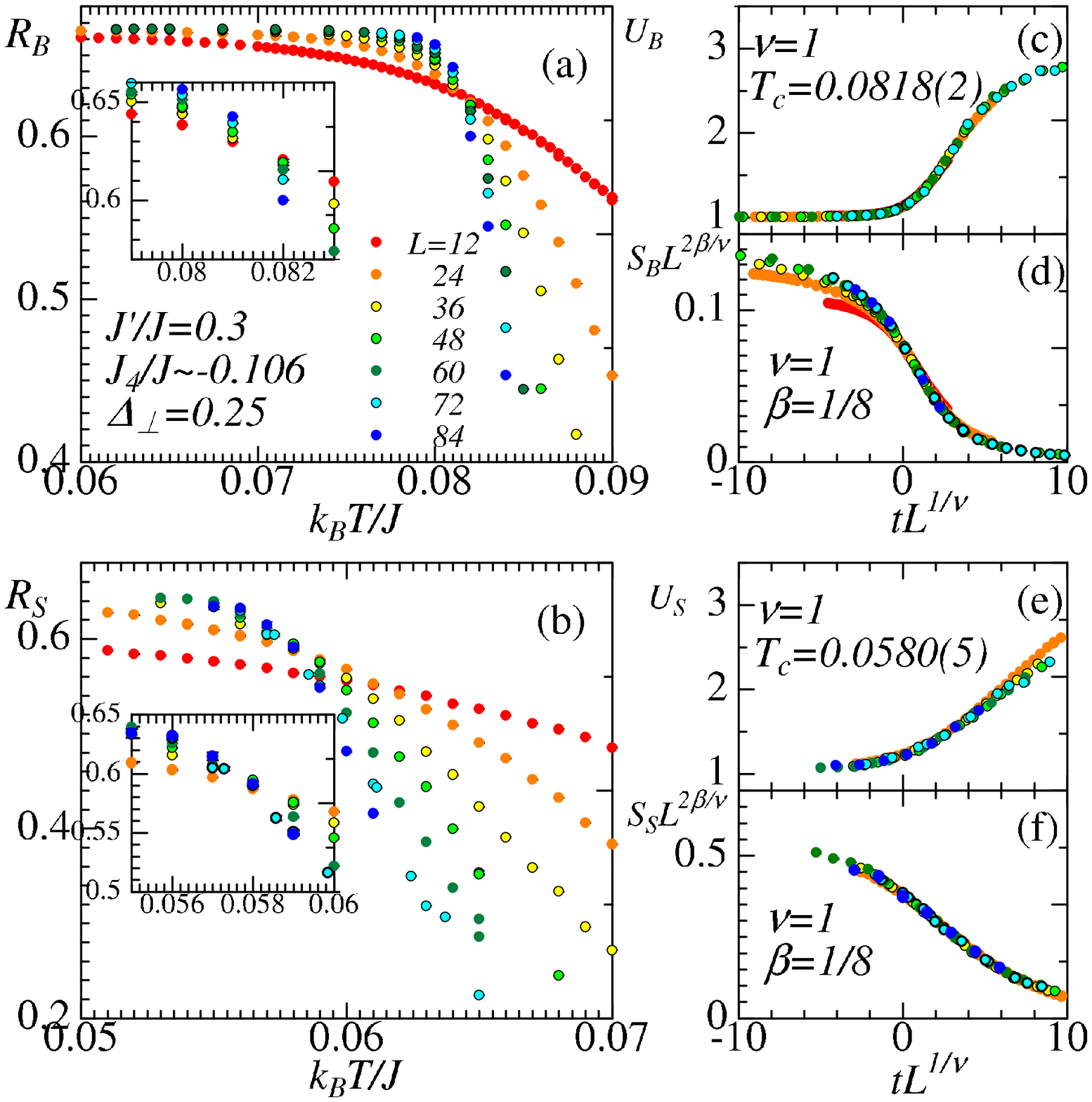}
  \end{center}
  \vspace*{-7mm}
  \caption{(color online) Finite-temperature transition to the $1/2$ plateau phase. (a) and (b) are temperature dependence of the Binder ratio of $R_B$ and $R_S$, respectively. (c) and (d) are the results of the finite-size scaling analysis for fourth-order cumulant $U_R$ of staggered magnetization $B_{st}$ and the static structure factor $S_B$ of bond spins. (e) and (f) correspond to (c) and (d) for $m_x^c$. In (f), the static structure factor of $m_x^c$ is calculated  from $S_S=\sum_{\bf r} m_x^c({\bf r})\exp[-i(\pi r_x +\pi r_y)]$}
\label{FSS_second}
\end{figure}

Next, we show the results of the quantum spin model with $\Delta_\perp=0.25$. We performed computations up to $L=84$.
Fig. \ref{FSS_second} shows the temperature dependence of the same order parameters as 
those defined in the above. The obtained results clearly indicate that the 
transition is a two-step process.

The curves for $R_B$ cross at $k_BT_{c1}/J=0.0815(3)$. This means that 
N\'eel order of the bond spins $B$ is realized for $T<T_{c1}$. Since a 
quotient group $C_4/C_2$, which is isomorphic to the group $C_2$, breaks down at $T_{c1}$, 
it is naively expected that the transition belongs to the two-dimensional 
Ising model universality class. Accordingly we performed 
finite-size scaling analysis assuming the scaling form $U_R=F(tL^{1/\nu})$ and $\nu=1$. 
We emphasize here that the data collapse shown in  Fig. \ref{FSS_second} 
(c) and (d) was obtained without any adjustable parameter, confirming that
the transition at $T_{c1}$ indeed belongs to the two-dimensional Ising universality 
class.

Fig. \ref{FSS_second} (b) shows the temperature dependence of the Binder ratio $R_S$ for 
$m_x^c$, and it provides an evidence of the other phase transition. The 
curves for $R_S$ for different system sizes intersect at $k_BT_{c2}/J=0.0580(5)$ 
and the difference $T_{c1}-T_{c2}$ is approximately  $0.03J$, 
slightly larger than the previously obtained value\cite{Suzuki}. The universality class 
of the phase transition at $T_{c2}$ should be the same as that of the two-dimensional 
Ising model because the remaining symmetry $C_2$ is also isomorphic to $Z_2$. 
This is confirmed by the finite-size scaling results presented in
Fig. \ref{FSS_second} (e) where we have used the same scaling function and critical 
exponent as for $R_B$, viz. $U_S=F(tL^{1/\nu})$ and $\nu=1$. Consequently, we conclude 
that there $exists$ an intermediate phase with  $C_2$-rotation symmetry phase that 
can be characterized by the bond 
N\'eel order accompanying the internal antiferromagnetic bare-spin fluctuation. The 
schematic spin configuration in this $C_2$-rotation symmetry phase 
is shown in Fig. \ref{spinconfig} (e).

\begin{figure}[htbp]
  \begin{center}
  \vspace*{2mm}
  \includegraphics[scale=0.5]{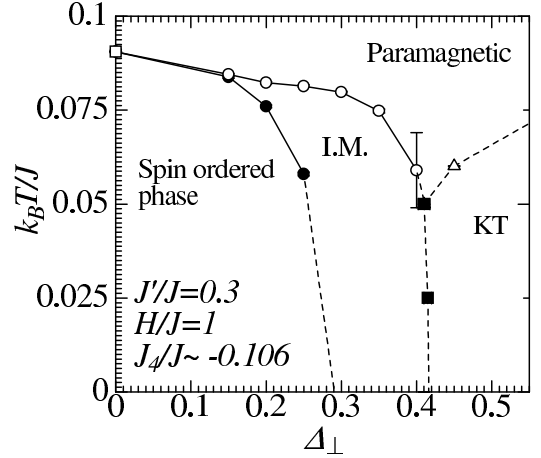}
  \end{center}
  \vspace*{-7mm}
  \caption{Phase diagram at $J'/J=0.3$ and $J_4/J=-(1/\sqrt{2})^3 \sim -0.106$. I.M. in the phase diagram means the $C_2$-rotational symmetric phase. Schismatic spin configurations in the spin ordered phase and I.M. are shown in Fig. \ref{spinconfig} (d) and (e), respectively. Open and solid circles indicate the second order transition with the two-dimensional Ising universality, and an open square is that with the four-state Potts universality. An open triangle is the Kosterslitz-Thouless transition. All lines are guide to the eyes.}
\label{phasediagram}
\end{figure}

If the antiferromagnetic fluctuation on the bond $J$ is relevant to the stabilization 
of the intermediate phase, the temperature range of the $C_2$-rotation 
symmetric phase expands upon increasing $\Delta_\perp$. In 
Fig.~\ref{phasediagram}, $\Delta_\perp$ dependence of the critical temperatures is 
presented for fixed $J'/J=0.3$, $H/J=1$ and $J_4/J=-(1/\sqrt{2})^3 \times J'/J 
\sim -0.106$. The results show that the critical temperature associated 
with the breaking of the $180\,^{\circ}$ rotational symmetry shifts to 
the lower temperature as $\Delta_{\perp}$ increases. This suggests that the 
dimerization of antiparallel spin on the diagonal bonds plays a key role
 in the transition. As discussed in the followings, the phase diagram is 
qualitatively understood from a comparison with that of the generalized 
chiral four-state clock model.

\section{Comparison to a classical model}
In the previous section, we have investigated the finite-temperature phase transition 
to the $m=1/2$ plateau phase for the classical and quantum spin models. Since the symmetry 
of both models is same and the finite-temperature phase transition is focused on, it 
is expected that the schematic phase diagram should be the same. However, the 
 results obtained are different from this expectation. In order to understand the discrepancy, 
we consider the generalized chiral four-state clock model. As we show below, 
this simplified classical model captures the characteristics of phase diagram 
for the original model.

We begin by introducing eight-spin unit clusters (windmill clusters) in the $m=1/2$ 
plateau state as shown in Fig. \ref{orderparameter}(b). 
In this phase, a down spin occupies one of four sites on a plaquette without 
the diagonal bond and up spins occupy the remaining three sites. 
If a down spin occupies the site labeled by `2' in Fig. \ref{orderparameter}(b), 
the nearest down spin prefers to locate at the third-neighbor distance from it as 
long as the ferromagnetic $J_4$ coupling. This indicates that the down spins in the 
$m=1/2$ plateau state occupies the same edge of diagonal bond connecting each plaquette. 
Since the ferromagnetic $J_4$ coupling is essential for stabilizing the 
$m=1/2$ plateau state (as discussed in Sec. II), it is reasonable to describe the 
$m=1/2$ plateau state by an arrangement of such clusters. This description becomes exact
in the limit $|J_4/J| \gg 1$ and $\Delta_{\perp}=0$.

\begin{figure}[htbp]
  \begin{center}
  \vspace*{2mm}
  \includegraphics[scale=0.45]{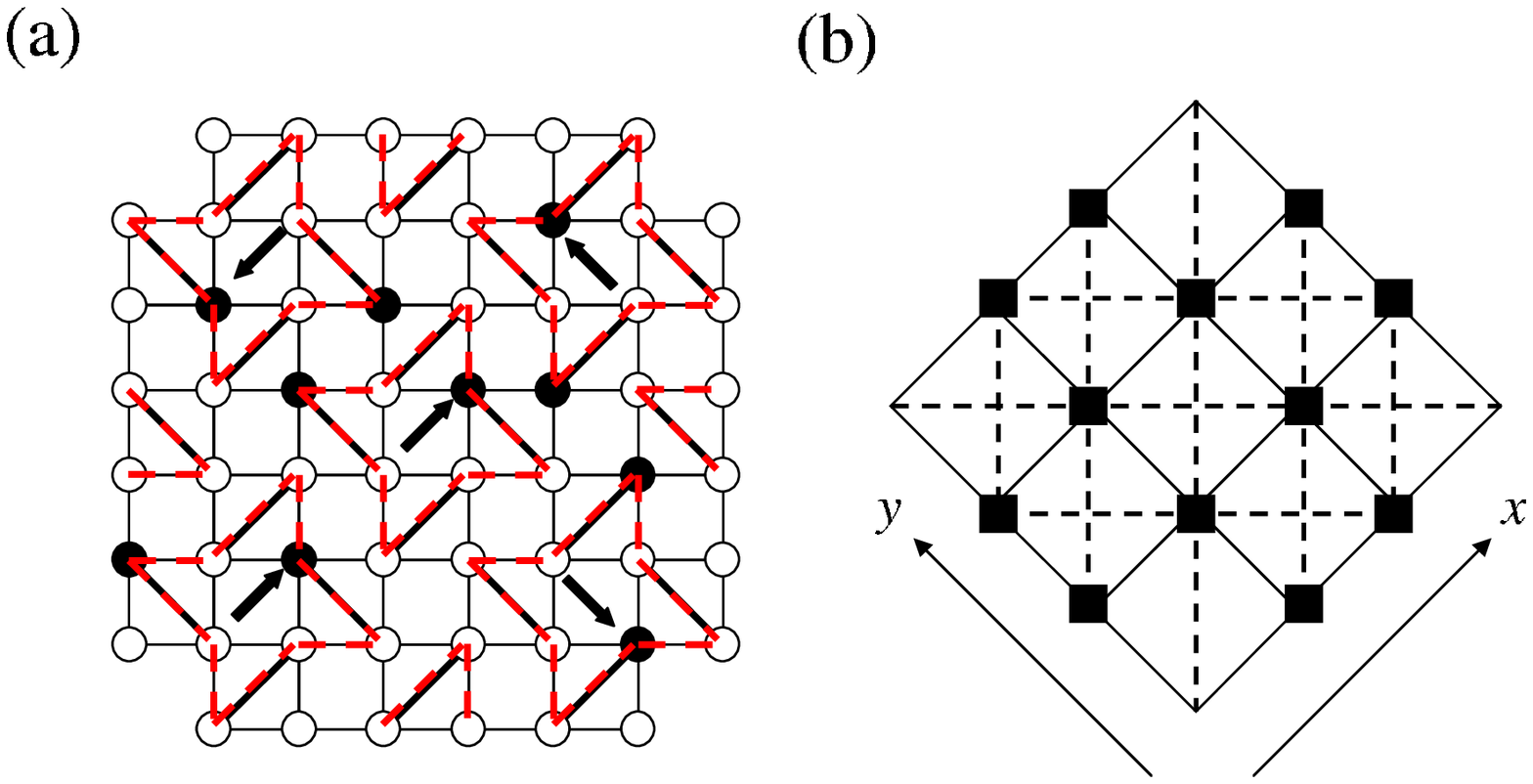}
  \end{center}
  \vspace*{-7mm}
  \caption{(a) Windmill clusters and clock spins in the simplified classical model (color online). A cluster is constructed by eight spins. Red dashed lines drawing windmills correspond to the clusters and arrows in the windmills denote angles of clock spins that represent states of the clusters. (b) Generalized four-state clock model on a square lattice. Solid squares correspond to each clusters described in (a). Solid (dashed) lines denote the nearest (next-nearest) neighbor interaction ${\mathcal J}_x$ and ${\mathcal J}_y$ (${\mathcal J}'_x$ and ${\mathcal J}'_y$). }
\label{ATmodel}
\end{figure}

In the Ising limit $\Delta_{\perp}=0$, 
the spin configuration in the $m=1/2$ plateau state can be expressed by the 
arrangement of the windmill clusters shown in Fig. \ref{ATmodel} (a). There are four kinds of 
clusters corresponding to the $90\,^{\circ}$-rotation symmetry breaking in the $m=1/2$ plateau. 
We assign clock spins pointing at an angle to each clusters. Here we ignore clusters having 
$m \ne 1/2$ for simplicity. Thus we obtain the simplified classical Hamiltonian that can 
be regarded as the generalized four-state clock model on a square lattice with the nearest 
and next-nearest neighbor interactions,
\begin{eqnarray}
{\mathcal H}_{sim}&=&{\mathcal H}_{\rm N.N.}+{\mathcal H}_{\rm N.N.N}\nonumber\\
{\mathcal H}_{\rm N.N.}&=&\sum_{i} \mathcal{J}_x(\theta_{i,j},\theta_{i+1,j})+\sum_{j} \mathcal{J}_y(\theta_{i,j},\theta_{i,j+1})\nonumber\\
{\mathcal H}_{\rm N.N.N}&=&\sum_{i} \mathcal{J'}_x(\theta_{i,j},\theta_{i+1,j+1})+\sum_{j} \mathcal{J'}_y(\theta_{i,j},\theta_{i-1,j+1}),
\label{GenClock}
\end{eqnarray}
where $\theta_{i,j}$ denotes an angle of clock spins and takes $0,\pi/2,\pi$, or $3\pi/2$. 
The value of $\theta_{i,j}$ denotes the position of a down spin in the windmill clusters (see Fig. \ref{ATmodel} (a)). 
${\mathcal J}_x$ and ${\mathcal J}_y$ are interactions between two clock spins, and 
their values are listed in Table I. The interactions, 
${\mathcal J}_x$ and ${\mathcal J}_y$, are calculated from the sums of coupling energy 
between spins located in different clusters. When $\Delta_{\perp}=0$, the values, A, B, C, 
and D, in ${\mathcal J}_x$ and ${\mathcal J}_y$ (see Table I) are evaluated as A=4$J_4$, B=2$J'$, C=0, 
D=4$J'$, respectively.  The resulting classical Hamiltonian retains $C_4$ 
rotational symmetry conditionally:  the system requires a lattice rotation 
when a global angle shift of all clock spins is executed. Such requirement arises from the 
geometric characteristics of the SSL. Since the symmetry of the Hamiltonian 
${\mathcal H}_{\rm N.N.}$ is lower than that of ${\mathcal H}_{\rm N.N.N.}$, we focus on 
the critical properties of ${\mathcal H}_{\rm N.N.}$ in the followings. 

\begin{table}[htbp]
\label{table1}
\caption{Coupling constants of the simplified classical Hamiltonian. Note that (A, B, C, D)=($4J_4$,$2J'$,$0$, $4J'$) corresponds to the parameter set we have treated in the section III. Upper (lower) side tables are matrices of the nearest neighbor (next-nearest neighbor) interaction. }
%
\begin{tabular}{lcccc}
\multicolumn{5}{c}{ ${\mathcal J}_x(\theta_{i,j},\theta_{i+1,j})$ } \\
$\theta_{i,j}\backslash\theta_{i',j'}$ & 0 & $\pi/2$ & $\pi$ & $3\pi/2$ \\
\hline
0 & A & B & C & B\\
$\pi/2$ & B & A & B & D\\
$\pi$ & C & B & A & B\\
$3\pi/2$ & B & -D & B & A\\
\hline
\end{tabular}
%
\begin{tabular}{lcccc}
\multicolumn{5}{c}{ ${\mathcal J}_y(\theta_{i,j},\theta_{i,j+1})$ } \\
$\theta_{i,j}\backslash\theta_{i',j'}$ & 0 & $\pi/2$ & $\pi$ & $3\pi/2$ \\
\hline
0 & A & B & D & B\\
$\pi/2$ & B & A & B & C\\
$\pi$ & -D & B & A & B\\
$3\pi/2$ & B & C & B & A \\
\hline
\end{tabular}
%
\hspace{15mm}
%
\begin{tabular}{lcccc}
\multicolumn{5}{c}{ ${\mathcal J}'_x(\theta_{i,j},\theta_{i+1,j+1})$ } \\
$\theta_{i,j}\backslash\theta_{i',j'}$ & 0 & $\pi/2$ & $\pi$ & $3\pi/2$ \\
\hline
0 & A & C & A & C\\
$\pi/2$ & C & A & C & -A\\
$\pi$ & A & C & A & C\\
$3\pi/2$ & C & -A & C & A\\
\hline
\end{tabular}
%
\begin{tabular}{lcccc}
\multicolumn{5}{c}{ ${\mathcal J}'_y(\theta_{i,j}\backslash\theta_{i-1,j+1})$ } \\
$\theta_{i,j}\backslash\theta_{i',j'}$ & 0 & $\pi/2$ & $\pi$ & $3\pi/2$ \\
\hline
0 & A & C & -A & C\\
$\pi/2$ & C & A & C & A\\
$\pi$ & -A & C & A & C\\
$3\pi/2$ & C & A & C & A \\
\hline
\end{tabular}
%
\end{table}

\begin{figure}[htbp]
  \begin{center}
  \vspace*{2mm}
  \includegraphics[scale=0.5]{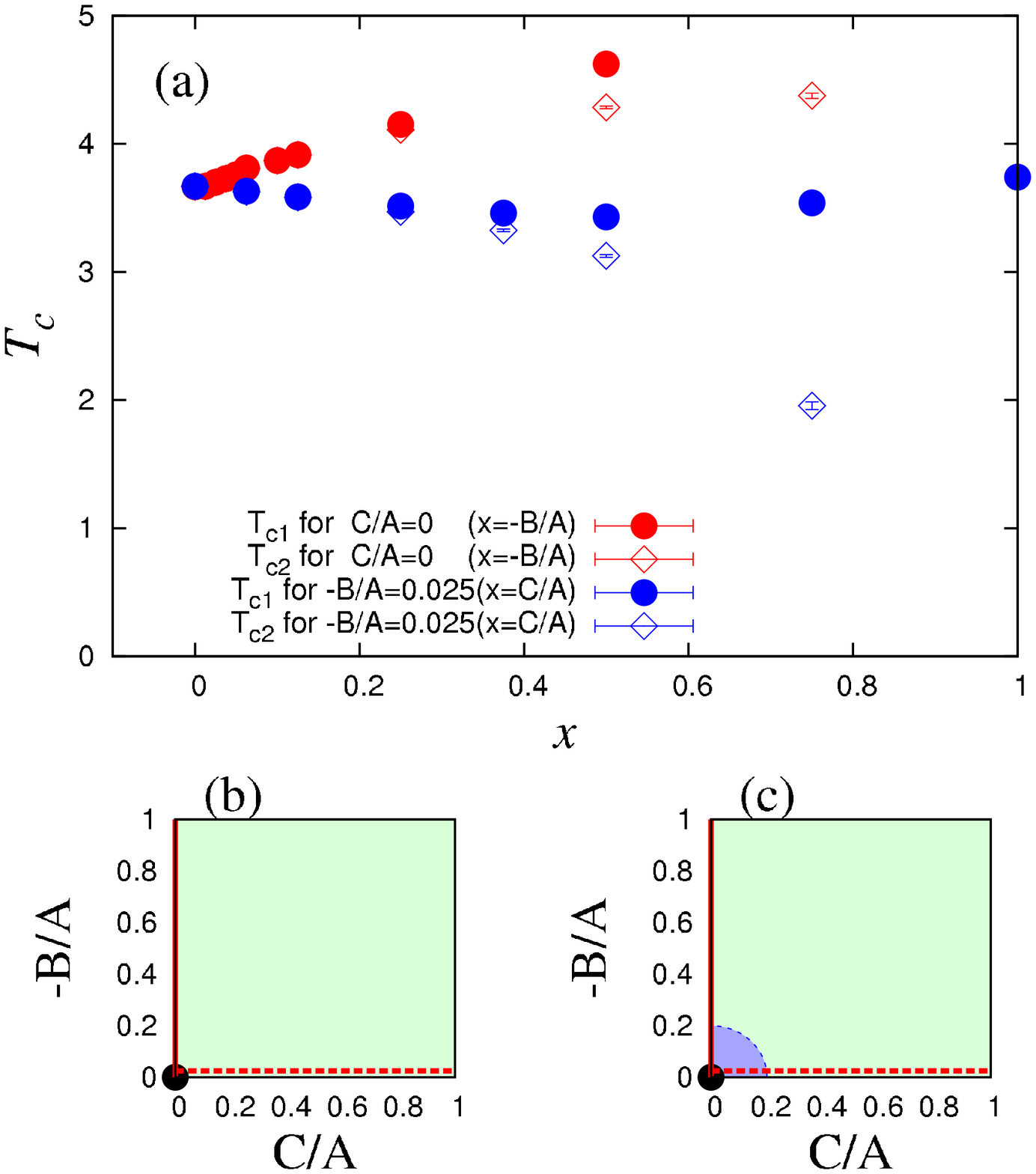}
  \end{center}
  \vspace*{-7mm}
  \caption{(color online) (a) Critical temperatures of the generalized four-state clock model when we fix at B/A=-0.05 and C/A=0. The higher and lower critical temperatures were calculated by the finite-size scaling analysis for the Binder ratio of $\langle \cos \theta \rangle$ and $\langle \cos 2\theta \rangle$, respectively. Red and blue symbols correspond to the cross sections on red solid and dashed lines drawn in (b) and (c), respectively. (b) and (c) are expected phase diagrams of the present clock model. In (b) and (c), an area painted by green indicates the region where the two-step phase transition with two-dimensional Ising universality takes place. A blue quarter circle in (c) means the area where the system undergoes a single phase transition.}
\label{Tcs}
\end{figure}

The simplified Hamiltonian ${\mathcal H}_{\rm N.N.}$ is identical to the 
conventional four-state Potts model in the limit B=C=D=0. Therefore, it is trivial that 
the finite-temperature phase transition at this point is a single second-order transition 
belonging to the four-state Potts universality. However, the universality 
class becomes nontrivial, when B$\ne$D$\ne$0. Performing Monte Carlo simulations, 
we numerically investigated the finite-temperature transition of the 
classical Hamiltonian ${\mathcal H}_{\rm N.N.}$ with D=2B and A=-4 up to $L=72$. We summarize the 
obtained critical temperatures in Fig. \ref{Tcs} (a). These critical temperatures were 
estimated from the finite-size scaling analysis for the Binder ratio and correlation 
ratio of the order parameters, $\langle \cos \theta \rangle$ and $\langle \cos 2\theta \rangle$.

\begin{figure}[htbp]
  \begin{center}
  \vspace*{2mm}
  \includegraphics[scale=0.5]{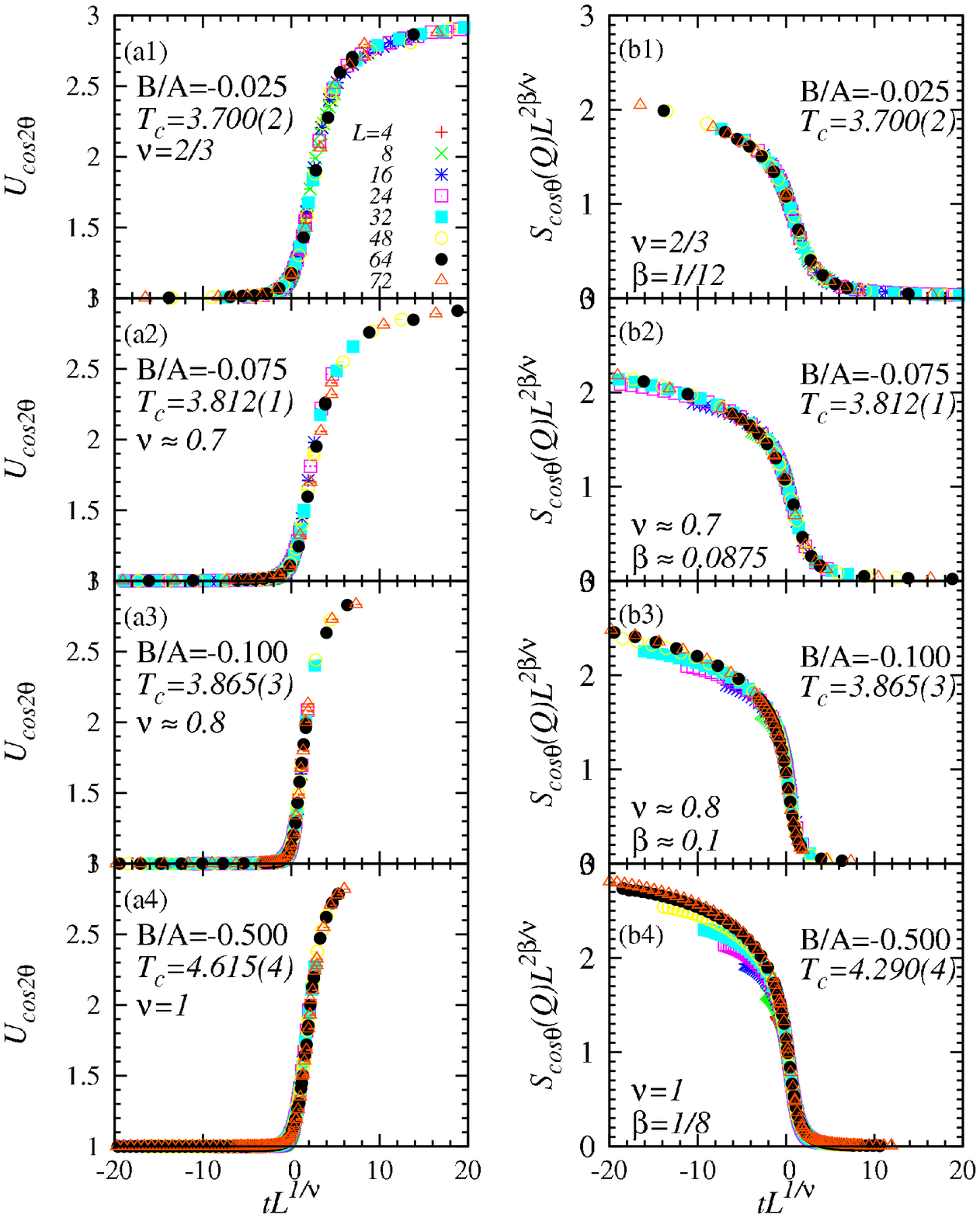}
  \end{center}
  \vspace*{-7mm}
  \caption{(color online) (a1) - (a4) ((b1) - (b4)) are the results of finite size scaling analysis for the fourth-order cumulant of $\langle \cos 2\theta \rangle$ (the static structure factor of $\cos \theta$) when C/A=0. In these analysis, we estimated all critical temperatures from crossing points of the correlation ratios $\langle \Gamma_{\cos \theta}(L/2)\rangle / \langle \Gamma_{\cos \theta}(L/4) \rangle$ and $\langle \Gamma_{\cos 2\theta}(L/2)\rangle / \langle \Gamma_{\cos 2\theta}(L/4) \rangle$. The data collapse in (a1) and (b1) ((a4) and (b4)) was obtained by using the fixed critical exponents $\nu=2/3$ and $\beta=1/12$ ($\nu=1$ and $\beta=1/8$). The critical exponents in (a2), (b2), (a3), and (b3) were estimated to minimize the discrepancies among data.}
\label{Clock40}
\end{figure}

In the following, we focus on the results when C/A=0 and B/A are 
varied (along the y-axis in Fig. \ref{Tcs} (b) and (c)). For C/A=0, the system 
undergoes a single second-order transition when the coupling ratio satisfies 
${\rm B/A} > -0.075$. To identify the universality class, we performed 
finite-size scaling analysis and present the results in Fig. \ref{Clock40}. 
(The scaling forms assumed here are the same as those shown in the previous section.) In 
the analysis, we estimated $T_c$ from the crossing of the curves for 
the correlation ratio, $\Gamma_{\cos n \theta}(L/2)/\Gamma_{\cos n \theta}(L/4)$, 
for different system sizes, where $\Gamma_{\cos n \theta}(r) = 
\langle \cos n \theta_{0,0} \cdot \cos n \theta_{r,r} \rangle$ and $n=1,2$.
For ${\rm B/A} > -0.07$, excellent data collapse is obtained with 
$\nu=2/3$ and $\beta=1/12$. This strongly suggests that the transition belongs 
to universality class of the four-state Potts model\cite{Domb}. For ${\rm B/A} < -0.075$, 
the critical exponent increases from $\nu=2/3$ to $\nu=1$ with 
the fixed ratio $\beta / \nu \sim 1/8$ as B/A decreases. For ${\rm B/A} \le -0.5$, we confirm 
a clear evidence of two-step phase transition with $\nu=1$ at both critical 
temperatures. Therefore, the critical properties for ${\rm B/A} \le -0.5$ belong 
to the two-dimensional Ising universality class.

\begin{figure}[htb]
  \begin{center}
  \vspace*{2mm}
  \includegraphics[scale=0.35,angle=270]{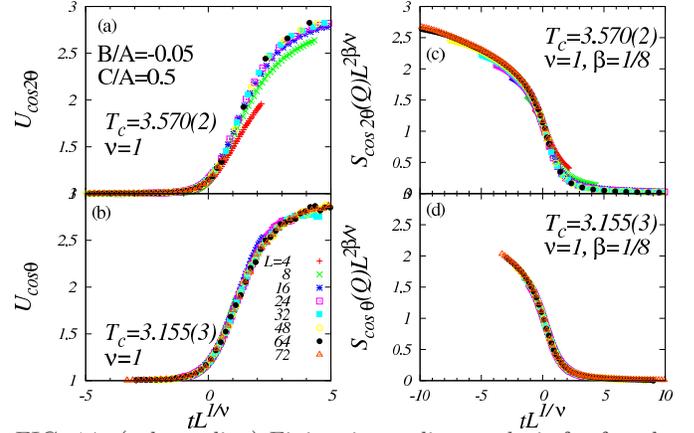}
  \end{center}
  \vspace*{-7mm}
  \caption{(color online) Finite size scaling analysis for fourth order cumulant of the order parameters $\langle \cos 2\theta \rangle$ (a) and $\langle \cos \theta \rangle $ (b) at B/A=-0.1 and C/A=0.5. (c) and (d) are the same analysis for the structure factor of each order parameter. In these analysis, each critical temperature is estimated from a cross of the correlation ratios $\langle 2\cos \theta_{L/2,0}\rangle/\langle \cos 2\theta_{L/4,0}\rangle$ and $\langle \cos \theta_{L/2,0}\rangle/\langle \cos \theta_{L/4,0}\rangle$. All results are obtained by using the fixed critical exponents $\nu=1$ and $\beta=1/8$.}
\label{Clock41}
\end{figure}

Next we discuss the results when the parameters are varied along the 
dashed lines in Fig. \ref{Tcs} (b) and (c), where B/A=-0.05 and $0 \leq {\rm C/A} 
\leq 1$. For ${\rm C/A>0.25}$, the system clearly undergoes a two-step phase transition with 
both transitions belonging to the two-dimensional Ising universality class. 
Fig. \ref{Clock41} presents the results of the finite-size scaling analysis at B/A=-0.05 and 
C/A=0.5. When the two-step phase transition takes place, the $90\,^{\circ}$ rotation symmetry breaks 
at the higher critical temperature and the $180\,^{\circ}$ rotation symmetry survives in the intermediate 
phase. The lower critical temperature decreases as the difference between A and C decreases. For 
${\rm C/A>1}$, the system seems to undergo a single phase transition and the $180\,^{\circ}$ rotation symmetry survives for any finite temperature.

The obtained results allow us to consider two scenarios of the phase diagram for the present four-state clock model. 
One is that the system shows a single phase transition only at B=C=0 and the two-step phase transition takes place for ${\rm B/A}<0$ and ${\rm C/A}>0$ as shown in Fig. \ref{Tcs} (b). 
The other is that there exists a region having the four-state Potts universality class around ${\rm B/A}={\rm C/A}=0$, denoted by the area in Fig. \ref{Tcs} (c). The critical exponents may show a crossover behavior at the boundary of the area. Further studies of the critical behavior of the extended four-state clock model are currently underway.

We compare the phase diagrams shown in Fig. \ref{phasediagram} and Fig. \ref{Tcs} here.
The topology of the phase diagram for the simplified classical model 
is the same as that of the Ising-like XXZ model on the SSL. In both models, the region of 
the intermediate phase, where only the $90\,^{\circ}$ rotational symmetry is broken, shrinks rapidly as 
the system approaches the four-state-Potts universality point. When we consider the quantum effects 
in the original Hamiltonian (\ref{Ham}), fluctuation of antiferromagnetic spin pairs on the diagonal 
bond is enhanced when $\Delta_\perp$ increases. Then, the sublattice magnetization of antiparallel 
spin pair closes to zero. In the case, where the antiparallel spin pair becomes a classical or 
quantum-triplet dimer with $S^z=0$, the clock spins pointing towards $\theta=0$ and $\pi$ 
(or $\pi/2$ and $3\pi/2$) direction can not be distinguished from each other. Such dimerization effect 
is captured as a reduction in the difference between A and C. Consequently, the lower transition 
temperature $T_{c2}$ associated with the $180\,^{\circ}$ rotational symmetry breaking shifts toward 
$T=0$ as $\Delta_\perp$ increases.

By considering the parameter set associated with the Ising limit of the original Hamiltonian, we 
can estimate a parameter set of ${\mathcal H}_{\rm N.N.}$ as A$\approx 4J_4$, 
B$\approx 2J'$, C$\approx 0$, and D$\approx 4J'$. For $J'/J=0.3$ and $J_4/J\sim -0.106$, 
${\mathcal H}_{\rm N.N.}$ shows a single phase transition with the critical exponent $\nu\sim 1$, 
in clear disagreement with that of the original model. This is because
we ignored the effect of the next-nearest neighbor couplings and the other states excluded from 
the four states treated here. A more quantitative estimation of the parameter set (A,B,C,D) 
is required by adding such effects for a quantitative mapping of the 
original Hamiltonian over the entire parameter range,  but this is beyond the scope of this paper.

\section{${\rm TmB_4}$}

\begin{figure}[htbp]
  \begin{center}
  \vspace*{2mm}
  \includegraphics[scale=0.475]{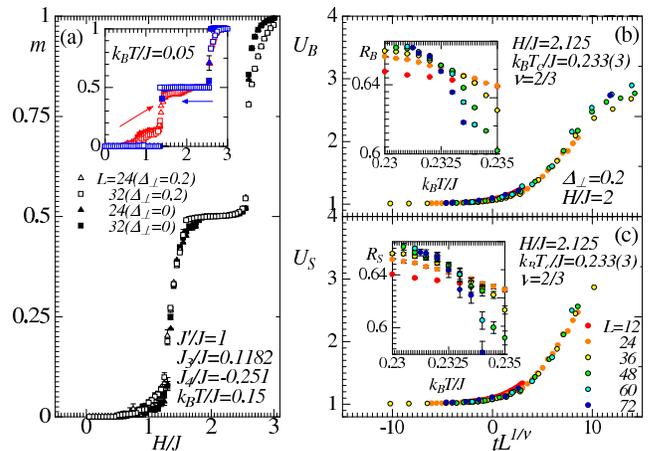}
  \end{center}
  \vspace*{-7mm}
  \caption{ (color online) (a) Magnetization curves at $J'/J=1$, $J_3/J=0.1182$ $J_4/J=-0.251$. Red solid (open) symbols represent the results of the Ising limit case (the quantum spin model case) when they have been obtained by the field sweeps from $H/J=3$ and the slow cooling. Blue ones are the results of the field sweeps from $H/J=0$. Inset of (a) is the results of short time Monte Carlo calculations when the field sweep is performed.  (b) and (c) show a finite-size scaling analysis for the fourth-order cumulant of $R_B$ and $R_S$, respectively. }
\label{Magcurve}
\end{figure}

Finally, we discuss the magnetization properties of ${\rm TmB_4}$. The $m=1/2$ plateau state is
observed for $H_{c1}\sim$1.9[T]$<H<$$H_{c2}\sim$3.6[T] at 2[K]\cite{TmB4_1}. 
The coupling ratio $J'/J \sim 1$ and a strong Ising anisotropy $\Delta_{\perp}<<1$ are
 expected from the crystal structure of ${\rm TmB_4}$ and 
specific heat measurements\cite{Siemensmeyer}. These experimental results 
help us estimate the other parameters for the present model. 
We obtain $J_3/J\sim0.1182$ and $J_4/J<-0.25$ at $J'/J=1$ from the local energy estimation 
in the Ising limit. 
The magnetization curves at fixed $J_3/J=0.1182$, $J_4/J=-0.251$, $J'/J=1$ and $k_BT/J=0.15$ are 
shown in Fig. \ref{Magcurve} (a). It is found that the magnetization jumps appear at 
$H_{c1}/J\sim 1.4$ and $H_{c2}/J\sim 2.65$ not only in the Ising limit 
but also for $\Delta > 0$. 
From these critical fields, we roughly estimate $(H_{c2}-H_{c1})$/$H_{c1} \sim 0.9$. 
This value is in good agreement with the experimental value. 
Therefore our estimation for the coupling ratio is quantitatively consistent 
with the critical fields of the magnetization jumps observed in ${\rm TmB_4}$.

The $m\sim 1/8$ plateau observed in the experiments \cite{TmB4_2,TmB4_4} appears to depend on the 
history of the system; $m\sim 1/8$ was observed around $H\sim 1.8$[T], when the fields were decreased 
from the saturation fields, whereas the magnetization remained vanishingly small around
$H\sim 1.8$[T] during the upsweep. 
In other words, a hysteresis loop was observed around $H\sim 1.8$[T]. 
The inset of Fig. \ref{Magcurve} (a) is the magnetization curves at $k_BT/J=0.05$ 
when we perform field sweeps. We started with the N\'eel state 
(the fully-polarized state) and increased (decreased) the applied field from $H/J=0$ ($H/J=3$). 
A hysteresis behavior similar to the experiments can be observed in short-time simulations. 
Although the Monte Carlo dynamics is not directly comparable to the real dynamics, this similarity is suggestive. 
The magnetization curves in the inset suggest a presence of shoulder around $m=1/8$ for $0.9<H/J<1.2$. 
In this field region, we have confirmed the existence of a mixed state 
 comprised of the N\'eel order and domain walls constructed of fully-polarized spin 
chains from the snapshot of spin configuration. The period of the domain walls is fluctuating and 
the average of the period changes continuously as the field decreases. Thus, the shoulder around 
$m=1/8$ seems to be a transient process in successive plateaus having the magnetization $0<m<1/2$. 
Since the magnetization value of the shoulder shifts to the lower magnetization value as the system 
size increases, it seems to be smeared out in the thermodynamic limit.  
More quantitative discussion of the $m=$1/8 shoulder is desirable via comparison with the other experimental observations.

We comment the finite-temperature transition to the $m=1/2$ plateau phase. Using the above parameters, 
we have studied it for the Ising spin model and the quantum spin model numerically.  
Fig. \ref{Magcurve} (b) and (c) are the results of the scaling analysis, when the quantum spin model is considered. The results indicate that a single second-order transition takes place and the 
critical exponent equals $\nu\sim 2/3$. The obtained results indicate that the critical property of 
the finite-temperature transition to the $m=1/2$ plateau phase can be explained by that of the 
four-state Potts model. This critical behavior was also confirmed in the Ising limit case. Therefore, 
if our estimation for the coupling constants is correct, we expect the experimentally observed 
critical exponents in ${\rm TmB_4}$ to belong to the four-state Potts universality class.

The $m=1/2$ plateau has been observed in the other SSL compound ${\rm ErB_4}$\cite{ErB4}. The magnetic 
moments of ${\rm ErB_4}$ are derived from ${\rm Er^{3+}}$ ions and the magnitude equals $J=15/2$. In 
this compound, the strong Ising anisotropy perpendicular to the SSL plains has been suggested. When we 
focus on the magnetic properties at a very low temperature, the effective Hamiltonian of the magnetic 
properties can be also described by the present model in the Ising limit. Therefore we believe that 
our results help understand the magnetic order of the $m=1/2$ plateau observed in ${\rm ErB_4}$.

\section{Summary}
In summary, we have studied the magnetic properties of a model of interacting 
spins with Ising-like exchange anisotropy and longer range interactions on the SSL that 
captures the low-temperature magnetic properties of the rare-earth tetraboride, ${\rm TmB_4}$. 
We have focused on the finite-temperature transition to the $m=1/2$ plateau phase and investigated 
the universality class of the transition. In the Ising limit, the system shows a single second-order 
transition. The critical behavior is well explained by that of the four-state Potts model. When there exists quantum exchange interactions, it has been confirmed that there is the two-step phase transition with 
an intervening intermediate phase. Both the transitions are belonging to the two-dimensional Ising 
universality class.  From the finite-size scaling analysis, we have ascertained that 
the intermediate phase can be characterized by the "$180\,^{\circ}$-rotation symmetric" state retaining 
$\pi$-rotation symmetry on individual triplets with $m_z=0$. 
Since the symmetry of the quantum spin model is the same as that 
of the classical spin version, it is naively expected that the critical behavior of both models 
will also be the same. However, the obtained results have indicated that the phase diagrams for 
the finite-temperature transition are different. To understand the critical behavior of both the
models, we have proposed the simplified classical model, namely the generalized four-state chiral 
clock model. By performing the Monte Carlo computations for the simplified classical model, we 
have found that the universality class at the critical temperatures and the topology of the phase 
diagram are in agreement with those of the original models on the SSL. Finally, we have studied 
the magnetic properties of ${\rm TmB_4}$. From the Ising limit analysis, we have
estimated the parameters that can explain (qualitatively) the magnetization curves observed in 
experiments. From the short-time Monte Carlo simulations, we have suggested that the $m\sim 1/8$ 
plateau seems to be a metastable state. 
We have also investigated the finite-temperature transition to the $m=1/2$ plateau state. If our estimation for the coupling constants is correct, measurements of the critical exponents belonging to the universality class of the four-state Potts model are expected in experiment.

\section*{Acknowledgments} 
The present research subject was suggested by C. D. Batista and we would like to thank him. The computation in the present work is executed on computers at the Supercomputer Center, Institute for Solid State Physics, University of Tokyo. The present work is financially supported by Grant-in-Aid for Young Scientists (B) (21740245), Grant-in-Aid for Scientific Research (B) (19340109), Grant-in-Aid for Scientific Research on Priority Areas ``Novel States of Matter Induced by Frustration'' (19052004), and by Next Generation Supercomputing Project, Nanoscience Program, MEXT, Japan.


\begin{references}
\bibitem{Footnote} The energy gains of the $1/3$ plateau, $1/2$ plateau and the fully polarized state are $\Delta E_{1/3}=J_3/12+J_4/6$, $\Delta E_{1/2}=J_4/2$, and  $\Delta E_{F}=J_3/4+J_4/2$.
\end{references}

\end{document}